\documentclass[sigconf]{acmart}

\usepackage{booktabs,kotex} 
\usepackage[T1]{fontenc}

\setcopyright{rightsretained}

\acmConference[]{DLRS 2018}{October 6, 2018}{Vancouver, Canada}

\usepackage{enumitem}

\begin{document}
\title{Deep Content-User Embedding Model for Music Recommendation}

\author{Jongpil Lee}
\affiliation{%
  \institution{KAIST}
  \city{Daejeon}
  \state{Korea}
  \postcode{34141}
}
\email{richter@kaist.ac.kr}

\author{Kyungyun Lee}
\affiliation{%
  \institution{KAIST}
  \city{Daejeon}
  \state{Korea}
  \postcode{34141}
}
\email{kyungyun.lee@kaist.ac.kr}

\author{Jiyoung Park}
\affiliation{%
  \institution{NAVER corp.}
  \city{Seongnam}
  \state{Korea}
}
\email{j.y.park@navercorp.com}

\author{Jangyeon Park}
\affiliation{%
  \institution{NAVER corp.}
  \city{Seongnam}
  \state{Korea}
}
\email{jangyeon.park@navercorp.com}

\author{Juhan Nam}
\affiliation{%
  \institution{KAIST}
  \city{Daejeon}
  \state{Korea}
  \postcode{34141}
}
\email{juhannam@kaist.ac.kr}

\renewcommand{\shortauthors}{Lee et al.}

\begin{abstract}
Recently deep learning based recommendation systems have been actively explored to solve the cold-start problem using a hybrid approach. However, the majority of previous studies proposed a hybrid model where collaborative filtering and content-based filtering modules are independently trained. The end-to-end approach that takes different modality data as input and jointly trains the model can provide better optimization but it has not been fully explored yet. In this work, we propose deep content-user embedding model, a simple and intuitive architecture that combines the user-item interaction and music audio content. We evaluate the model on music recommendation and music auto-tagging tasks. The results show that the proposed model significantly outperforms the previous work. We also discuss various directions to improve the proposed model further. 
\end{abstract}

%
%



\maketitle

\section{Introduction}

Music recommendation has gained more attention in recent years as accessibility to music has dramatically increased by music streaming services and recommending songs that satisfy users' taste has become essential in the services. Music recommendation is also an important feature of AI speakers that are widely spreading recently. However, building a successful music recommendation system still remains as a challenging problem, because of the semantic gap between user context and music content \cite{celma2010music}. 


Common approaches to solve the music recommendation problem can be broadly divided into Collaborative Filtering (CF) and Content-Based Filtering (CBF) \cite{celma2010music,schedl2018current}. The CF approach, which utilizes user listening counts or song ratings, is an effective solution, but it suffers from the cold-start problem especially for new items. On the other hand, the CBF approach, which directly uses the content information (e.g. tag or audio), can solve the item cold-start problem, but it is difficult to exploit common consumption patterns from users or song popularity.      


To bridge this gap, hybrid approaches, which incorporate different modalities (e.g. user and content data) into one system, have been explored. A prominent approach is \textit{deep content-based music recommendation model} proposed by Oord \textit{et al.}, in which the CBF model is trained to predict the item latent factor obtained from the CF model \cite{van2013deep,platt2017content}. This was further extended to take not only audio content, but also artist biographies \cite{oramas2017deep}. On the contrary, the CF model can also be trained to match up with the pre-trained CBF model \cite{liang2015content}. However, these hybrid approaches still have a problem in that the learning stages of the CF and CBF models are separated, which can yield a sub-optimal solution.  


To address this problem, an end-to-end model, which can jointly learn from both content and user data, has been explored. Wang and Wang unified deep belief network and probabilistic graphical model for the audio and user models, respectively \cite{wang2014improving}. However, the audio model requires a pre-training stage or the model should be simplified to train the overall model. Chou \textit{et al.} used Convolutional Neural Networks (CNN) for the audio CBF model and multi-layer perceptron (MLP) for the preference vector \cite{chou2017conditional}. The preference vector, which is the survey or aggregated result of the artists and genres for each user, provides benefits in solving user cold-start problem, but it also has a limitation in that the music consumption pattern is simplified, potentially making the personalized recommendation less reliable. 


\begin{figure*}[!th]
\centering
 \includegraphics[width=\textwidth]{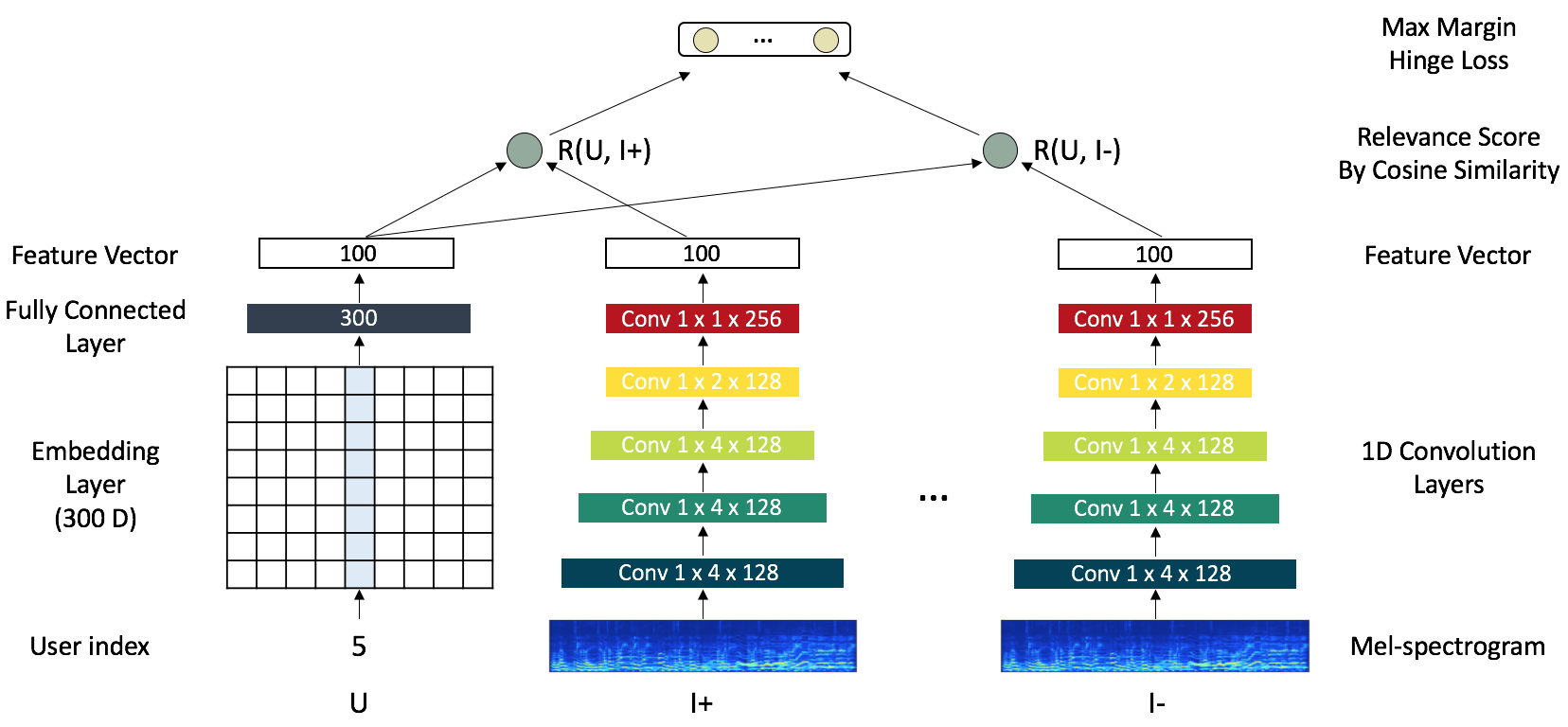}
  \caption{Deep content-user embedding model.}
  \label{fig:mainfig}
\end{figure*}


Meanwhile, Deep Learning (DL) based recommendation systems have been actively investigated in recent years due to the success of DL in many application domains \cite{hidasi2017dlrs,zhang2017deep}. Especially, Neural Collaborative Filtering (NCF) introduced the general framework to conduct CF using deep learning techniques \cite{he2017neural}. The overall architecture is as follows: For each user-item interaction, the user and item indexes are injected to each side of user and item embedding layer. Then, the user and item features are combined (this can be done using element-wise product or simple concatenation) and used to predict its interaction score with several layers of MLP. Another interesting approach is Deep Structured Semantic Model (DSSM) \cite{huang2013learning,elkahky2015multi}. The architecture of DSSM is more similar to the matrix factorization-based CF than the NCF, because the objective of the model is directly calculated from the similarity score of the user and item's high-level latent factors. This has the benefit of directly utilizing the latent factors for the user or item similarity-based search or filtering.

In this work, we introduce \textit{deep content-user embedding music recommendation model}, a hybrid DL-based music recommender model, which utilizes user-item interaction and song audio together in an end-to-end manner to solve the item cold-start problem. We evaluate the model for music recommendation and music auto-tagging tasks. The results show that the proposed model outperforms the baseline \cite{van2013deep} for both tasks. The advantage of the proposed model lies in the architecture and objective function, which provide a simple and intuitive way to handle the domain gap between the user-item interactions and unstructured audio data.


\section{Deep content-user embedding model} \label{section2}

The overall architecture that we propose is illustrated in Figure \ref{fig:mainfig}. Since the two different modalities, user-item interaction and audio (the content of item), are jointly trained in a single model, the following issues should be considered when configuring the model architecture.

\begin{itemize}
  \item The first layer configuration on the user side to handle a large number of users
  \item Method to combine the user and item feature vectors
  \item Loss function and training strategy
\end{itemize}

First, a huge number of users may lead to a computational burden when the raw user one-hot vector is utilized with the fully-connected layer in the first layer on the user side. To overcome this problem, the DSSM \cite{huang2013learning} used \textit{n}-gram based word hashing technique to reduce the vocabulary size in the first layer of their application. In \cite{chou2017conditional}, they used a  similar hashing technique, but using a preference vector instead. On the other hand, a lookup-table style embedding layer has been introduced in training word representations \cite{mikolov2013distributed}. This method has an advantage of saving the computational cost by updating only the selected user's embedding, while keeping the diversity of the user's taste profile. Therefore, we use the lookup-table style embedding layer for the first layer on the user side. 


Second, the method of combining the user and item factors can be varied. There are broadly two types of combining methods: feature fusion and pair-wise similarity calculation. The feature fusion method can be further divided into element-wise multiplication \cite{chou2017conditional} and concatenation \cite{oramas2017deep}. After the features are combined, fully-connected layers are often added to make a binary prediction. In this work, we explore the pair-wise similarity calculation following \cite{huang2013learning}. This is because the pair-wise calculation is more similar to the matrix factorization method that can provide high-level user and item factors. This can also be applied to the user and item similarity calculation or filtering. The relevance score calculation between the user feature vector and the item feature vector is another issue that we should consider. In our preliminary experiments, we compared cosine similarity score and dot product. The cosine similarity method was more effective in the proposed model. Thus, the relevance score between the user feature vector and the item feature vector is defined as the following cosine similarity score: 

\begin{equation} \label{eq:cosine}
R(U,I) = \cos(y_U,y_I) = \frac{y_U^Ty_I}{|y_U||y_I|}
\end{equation}
where $y_U$ and $y_I$ are the feature vectors of the user and item, respectively. 


Third, the choice of loss function and training strategy is also important. We apply the negative sampling technique over the pair-wise similarity score function following \cite{huang2013learning}. Therefore, for each user-item interaction, the model takes one listened song and several non-listened songs as the input on the item side. Then, the loss is calculated using the relevance scores obtained from the user feature vector and the item feature vectors. We tested two loss functions: One is the softmax function with categorical cross-entropy loss to maximize the positive relationships and the other is the max-margin hinge loss to set margins between positive and negative examples \cite{frome2013devise}. In our preliminary experiments, the proposed relevance score with negative sampling technique was successfully trained only with the max-margin loss function between the two objectives, which is defined as follows:

\begin{equation} \label{eq:cosine}
loss(U,I) = \sum\limits_{I^-}\max[0, \Delta - R(U,I^+) + R(U,I^-)]
\end{equation}
where $\Delta$ is the margin, $I^+$ and $I^-$ denote positive example and negative examples, respectively. We also grid-searched the number of negative samples and the margin, and settled on the number of negative samples as 20 and the margin value $\Delta$ as 0.2. The settings of the audio model and the training strategy are mostly adopted from \cite{park2018representation}. We note that the user play counts have been binarized in this work, following \cite{he2017neural}. In other words, when the user-item interaction occurs at least once, we treat them as a positive case, otherwise negative. We leave the more advanced loss function, which can take into account the raw implicit feedback data, as our future work.

Once the model is trained, the prediction of the user's preference on the non-listened song is simply calculated through the relevance score function of the user and item feature vectors, which is extracted from the user's index and the item's audio, respectively.

\section{Experimental Settings}
In this section, we describe the model setup and training details.

\subsection{Model Setup}
The proposed model consists of two networks, user model and audio model, as depicted in Figure \ref{fig:mainfig}. The user model is constructed using an embedding layer and fully-connected layers. The audio model is configured using one-dimensional convolution layers, sliding over only temporal dimension. Mel-spectrogram is used as an input to the audio model and the model parameters are shared across all positive and negative samples. The audio model is composed of 5 convolution and max pooling layers. For both user and audio model, rectified linear unit (ReLU) activation layers are used after every layer except for the feature vector layer. 

\subsection{Dataset}
The proposed model was evaluated in a similar settings to that in \cite{van2013deep}. We use the Million Song Dataset (MSD) \cite{bertin2011million} along with two associated datasets, Echo Nest Taste Profile Subset and Last.fm dataset. The Echo Nest Taste Profile Subset provides play count data for over 380,000 songs and one  million users. The Last.fm dataset offers tags for over 500,000 songs. We validate the model with two tasks, music recommendation and music auto-tagging. In the recommendation experiments, we use a subset of the Echo Nest Taste Profile Subset, which contains the most frequently listened 10000 songs and the most active 20000 users. In the auto-tagging experiments, we first filter the song list and tag list to have at least one of the 50 most-used tags. Then, we leave 6903 songs that are present in both the recommendation song subset and the tagging song list. In both tasks, we randomly split the song lists into train/valid/test sets at a proportion of 70\% /10\% /20\%. We use AUC (Area Under Receiver Operating Characteristic) as a primary evaluation metric for the both tasks.

\subsection{Training Details}
For the audio preprocessing, we compute the spectrogram using
1024 samples for FFT with a Hanning window, 512 samples for hop size and 22050 Hz as sampling rate. We then convert it to mel-spectrogram with 128 bins along with a log magnitude compression. We choose 3 seconds as a context window of the audio model input following previous work \cite{park2018representation}. Out of the several seconds audio, we randomly extract the context size audio and put them into the network as a single example. The input normalization is performed by dividing standard deviation after subtracting mean value across the training data. We optimize the loss using stochastic gradient descent with 0.9 Nesterov momentum with 1e−6 learning rate decay. Our system is implemented in Python 2.7, Keras 2.1.1 and Tensorflow-gpu 1.4.0 for the back-end of Keras \cite{chollet2015keras}. We use NVIDIA Tesla M40 GPU machines for training our models. 

\section{Evaluation}
In this section, we explain the baseline model and the evaluation methods.

\subsection{Task1 : Music Recommendation}
To validate the effectiveness of the proposed model, we reproduce the \textit{deep content-based music recommendation model} \cite{van2013deep} and compared them with our model. They used Weighted Matrix Factorization (WMF) algorithm for CF \cite{hu2008collaborative} and we implement it with the Implicit python library \cite{implicit}. For predicting latent factors from the corresponding audio, we use the same audio model as the audio side of the proposed model. We perform the evaluation by reconstructing the user-item matrices using the item latent factor obtained from WMF and also using the predicted latent factor from the corresponding audio. We term the former as \textit{WMF} and the latter as \textit{WMF+Regression}. Similarly, we reconstruct the user-item matrix using the relevance score between the user and item feature vector in our work. Finally, we calculate the AUC scores for each user and averaged them. We also report popularity-based recommendation result as a baseline of the recommendation performance. This means that it recommends the same songs to all users based on the total number of songs played.

\subsection{Task2 : Music Auto-Tagging}
To verify the proposed end-to-end architecture and the feature vector's usefulness, we also conduct music auto-tagging task in a transfer learning setting. The item factor, predicted item factor and item feature vector from the \textit{WMF}, \textit{WMF+Regression} and \textit{Deep Content-User Embedding Model} is used as a feature vector of a song. We then trained a 2-layer MLP on top of the feature to predict the tags.
\vspace{3mm}

\section{Results}
In this section, we examine the proposed model and compare
them to the reproduced results of \cite{van2013deep}.

\begin{table}[t]
	\caption{Music recommendation results.}
	\label{table1}
\begin{center}
\begin{tabular}{clc}
\toprule
Type & Models & AUC \\
\midrule
\textemdash & Popularity & 0.7059\\
CF & WMF  & 0.9302\\
Hybrid & WMF+Regression \cite{van2013deep}    & 0.6967  \\
Hybrid & Deep Content-User Embedding Model & 0.7914 \\
\bottomrule
\end{tabular}
\end{center}
\end{table}

\begin{table}[t]
	\caption{Music auto-tagging results.}
	\label{table2}
\begin{center}
\begin{tabular}{clc}
\toprule
Type & Models & AUC \\
\midrule
CF & WMF  & 0.8683\\
Hybrid & WMF+Regression \cite{van2013deep}   & 0.7876  \\
Hybrid & Deep Content-User Embedding Model & 0.8450 \\
\bottomrule
\end{tabular}
\end{center}
\end{table}
\vspace{2mm}

\subsection{Task1 : Music Recommendation}
Table \ref{table1} shows the recommendation results. We can first find that the reproduced result of the \textit{WMF+Regression} is in a similar performance level to the original work (0.70987). However, the score is comparable to the popularity-based recommendation result. Second, the result of the proposed model shows significantly better performance than it. The performance gain is seen to be from the end-to-end optimization architecture of the proposed model. Third, we can see that there still exists a large difference between the \textit{WMF} and the proposed model. To examine the bottleneck of the proposed model, we tested another model that has the same architecture as the proposed model on the user side, but audio model. In this model, the item side is constructed with the same architecture as the user side so the model takes user and item indexes rather than user and audio. This comparison showed 0.7832 score, which is close to the proposed model. We analyze this result as follows. Bridging the gap between user-item interaction and unstructured audio data can be solved using a proper deep learning based recommender system. Also, it seems that the choice of loss function and learning strategy affects the model performance rather than the choice of the input modality. 
\vspace{2mm}

\subsection{Task2 : Music Auto-Tagging}
From the Table \ref{table2}, we can also find that the reproduced result of \textit{WMF} is close to the previous work (0.86703). In the music auto-tagging task, the performance gap between the proposed model and \textit{WMF} seems to be small and the proposed model largely outperforms the \textit{WMF+Regression}. We believe that this result is obtained from the advantages of the proposed end-to-end architecture. 

\vspace{2mm}

\section{Conclusion and future work}
In this work, we presented the deep content-user embedding model to simultaneously learn the user-item interaction and unstructured audio data in an end-to-end fashion. The proposed model consists of the user and item sides, each of which takes user index and multiple audio as input, respectively. The model architecture and learning strategy are designed to preserve the advantage of the conventional matrix factorization based recommendation approach. The embedding layer in the user side enables the computational cost to be reduced. The item side takes audio directly to solve the item cold-start problem. The model is verified in the music recommendation and music auto-tagging tasks. 
The results showed that the proposed model is highly more effective than the \textit{deep content-based music recommendation model}. When comparing the results to the conventional matrix factorization method which has the cold-start problem, there is still a room to improve. However, we believe that the proposed model shows a good direction to utilize deep learning techniques to solve the recommendation problems.

For future work, we will thoroughly investigate the following issues. First, various loss function and learning strategy should be explored. As introduced in Section \ref{section2}, there exist many choices in constructing the architecture. For example, the way of conducting negative sampling can be varied depending on the choices of the loss function (pair-wise vs point-wise loss) and it can significantly affect the model performance. Second, we used binarized implicit feedback data in this work, which may affect the learning property of the model. Raw implicit feedback data can have more information about users' music consumption pattern or song popularity information. Third, the evaluation metric should be diversified to measure different aspects of the architecture. For example, AUC score and precision at K may have different trends if raw implicit feedback can be utilized to train the model due to its popularity bias property. Fourth, more advanced audio models such as \cite{kim2017sample} can be applied to the item side of the proposed model. Lastly, the model can be extended to have a multi-view architecture \cite{elkahky2015multi}. This method allows interaction between different domains (e.g. user-item interactions, audio, artist bibliographies and album images), which can make the feature representation richer. 

\vspace{2mm}

\begin{acks}
This work was supported by Basic Science Research Program through the National Research Foundation of Korea funded by the Ministry of  Science, ICT \& Future Planning (2015R1C1A1A02036962) and by NAVER Corp.
\end{acks}

\bibliographystyle{ACM-Reference-Format}
\bibliography{sample-bibliography}

\end{document}